\documentclass[%
 reprint,
nofootinbib,
 amsmath,amssymb,
 aps,
]{revtex4-1}

\usepackage{graphicx}% Include figure files
\usepackage{dcolumn}% Align table columns on decimal point
\usepackage{bm}% bold math
\begin{document}

\preprint{APS/123-QED}

\title{What is the dual of two entangled CFTs?}% Force line breaks with \\

\author{Samir D. Mathur}
 %\altaffiliation[Also at ]{Department of Physics,\\ The Ohio State University,\\ Columbus,
%OH 43210, USA}%Lines break automatically or can be forced with \\
 \email{mathur.16@osu.edu}
\affiliation{Department of Physics, The Ohio State University, Columbus,
OH 43210, USA
}%


\def\nn{\nonumber \\}
\def\p{\partial}
\def\t{\tilde}
\def\h{{1\over 2}}
\def\be{\begin{equation}}
\def\bea{\begin{eqnarray}}
\def\ee{\end{equation}}
\def\eea{\end{eqnarray}}
\def\b{\bigskip}


\begin{abstract}

It has been conjectured that the dual of the eternal black hole in AdS is two entangled but disconnected CFTs. We show that the entanglement created by the process of Hawking radiation creates several challenges for this conjecture. The nature of fuzzball states suggests a different picture, where the dual to two entangled CFTs is two entangled but disconnected spacetimes. We argue for a process of `quick tunneling' where the Einstein-Rosen bridge of the eternal hole tunnels rapidly into fuzzball states, preventing the existence of the eternal hole as a semiclassical spacetime. The regions behind the horizon  then emerge only in the approximation of fuzzball complementarity, where one considers the impact of probes with energy $E\gg T$. 

\end{abstract}

%Use showkeys class option if keyword
\maketitle

%\tableofcontents

\section{\label{secone}Introduction}

Hawking's discovery of black hole evaporation led to a deep puzzle \cite{hawking}. Particle pairs are created by the gravitational field  around the horizon. One member of the pair, $b$, escapes to infinity as radiation, while the other member $c$ falls into the hole to reduce its mass. These two particles are in an entangled state, so the entanglement of the  radiation with the remaining hole keeps rising. This leads to a puzzle near the endpoint of evaporation: how can the small residual hole have  the huge degeneracy required to carry this entanglement? 

Many aspects of string theory suggest that the evaporation of the hole should be no different from the burning away of a piece of paper; thus we should {\it not} have such a monotonically growing entanglement. In \cite{cern} it was shown, using strong subadditivity, that small corrections to the physics around the horizon cannot resolve the problem; one needs corrections of order {\it unity}.\footnote{See \cite{giddings,avery,acp} for furthur comments in this direction.} Then we have, a priori, two possibilities:

\b

(P1)   The black hole  has a  traditional horizon, where the spacetime around the horizon is in the local vacuum state. Then entangled pairs will be produced at the horizon, but one can conjecture that some new (nonlocal) effect solves the problem of growing entanglement. In discussing this possibility, we will focus on the recent proposal of Maldacena and Susskind \cite{maldasuss2} where it is conjectured that entangled particles are connected by a `wormhole', regardless of how far apart they are.

(P2) The black hole does not have a traditional horizon, so the radiation is not emitted through the Hawking process of pair creation. In this case we are not forced into Hawking's problem of rising entanglement. But the nontrivial task is to find the alteration of the state at the horizon, since the `no-hair' theorems suggest that the hole always settles down to its unique metric which has the vacuum state around the horizon. In string theory we find the fuzzball construction \cite{fuzzballs}, which evades the no-hair theorems \cite{gibbonswarner} and gives the required modification of the hole. 

\b

\begin{figure}[h]
\includegraphics[scale=.42]{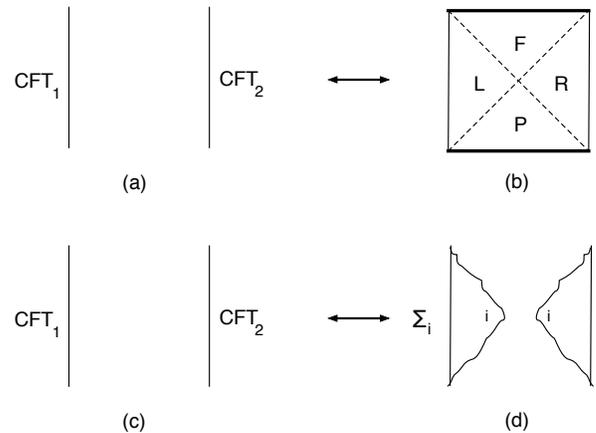}% Here is how to import EPS art
\caption{The conjecture of \cite{eternal} says that two entangled CFTs (a) gives the connected spacetime (b). The nature of fuzzballs suggests that two entangled CFTs (c) give two entangled but disconnected spacetimes (d). \label{e2} }
\end{figure}

A significant role in this debate has been played by consideration of the eternal hole in AdS space. We are interested in the notion of AdS/CFT duality in the context of this eternal hole. The eternal hole in AdS has {\it two} asymptotically AdS boundaries, so the usual notion of AdS/CFT duality \cite{maldacena} suggests that the eternal hole spacetime is dual to two CFTs. The two AdS boundaries are not connected, so we have two {\it disconnected} CFTs. Corresponding to the possibilities (P1), (P2) above, we have the following two possibilities:

\b

(P1')  In \cite{eternal}, it was conjectured that when these disconnected CFTs are placed in a particular entangled state, the CFT dual is the eternal hole. Thus when disconnected CFTs are placed in a state that is entangled, the dual spacetime can be {\it connected} (fig.\ref{e2}(a, b)). 

(P2') In the fuzzball proposal, each state of a CFT is dual to a spacetime that ends before reaching the horizon. This suggests that the dual to the entangled CFTs is just a pair of {\it disconnected} spacetimes, with wavefunctionals in the corresponding entangled state (fig.\ref{e2}(c),(d)). The spacetime fig.\ref{e2}(b) only emerges as an approximate description for quanta that impinge on the fuzzball surface with $E\gg T$, a notion termed `fuzzball complementarity' \cite{fcomp,beyond,mt1,mt2}. 

\b

In this paper we will observe that the picture (P1') faces three different (though related) challenges when confronted with the information puzzle.  For each such challenge, we will try to see if it is possible to find an escape route by requiring some appropriate behavior from the wormhole picture of \cite{maldasuss2}.   We will also note how the corresponding challenges can be addressed in the fuzzball situation (P2').

\section{\label{sectwo}The eternal hole and the wormhole conjecture}

In this section we recall details of the  proposal of \cite{eternal} depicted in fig.\ref{e2}(a),(b), and how it leads to the recent proposal that wormholes connect entangled systems \cite{maldasuss2}.

\subsection{\label{sectwo:1}The conjecture of \cite{eternal}}

Fig.\ref{e2}(b) depicts the Penrose diagram of the eternal hole in AdS.  There are {\it two} asymptotically AdS boundaries in this spacetime. By the normal expectations of AdS/CFT duality, one would argue that the eternal black hole spacetime is dual to a CFT that lives on the union of these two boundaries. But these two boundaries do not touch; thus the CFT defined on each boundary does not interact with the CFT on the other boundary. The only relation between the CFTs is that the  overall state is an entangled one \cite{eternal}
\be
|\Psi\rangle=\sum_i e^{-{E_i\over 2T}} |E_i\rangle_L |E_i\rangle_R
\label{one}
\ee
where $T$ is the temperature of the hole and $|E_i\rangle_L, |E_i\rangle_R$ are the energy eigenstates of the CFTs on the left and right sides of the diagram. Note that the dual gravity system - the eternal hole - has four spacetime regions separated by the horizons of the hole. In fig.\ref{e2}(b) we label these as R (right), L (left), F (future) and P (past). The spacetime is assumed to continue smoothly across the horizons, so these four regions together make a connected spacetime describing the full geometry of the eternal hole.

\subsection{\label{sectwo:2}The conjecture of \cite{maldasuss2}}

Suppose we accept the duality indicated in fig.\ref{e2}(a),(b). Then we observe that a pair of systems  (\ref{one}) that are entangled but {\it disconnected}  have a dual representation where the spacetime is {\it connected}. van Raamsdonk took this argument further by assuming a gravity dual $|g_i\rangle$ that is dual to each eigenstate $|E_i\rangle$ of the CFT.  From this he concluded that if we have two spacetimes that are  entangled but disconnected,  then we can represent them by a  spacetime which is {\it connected} \cite{raamsdonk}.  

Note that the  left and right sides of the eternal hole are joined by a {\it wormhole}, the `Einstein-Rosen  bridge'. It was conjectured in \cite{maldasuss2} that this occurrence should be elevated to a general principle: whenever two system are entangled, there should be some version of a physical spacetime connection between them. This  `wormhole'  could be quite `quantum' for systems with a small number of degrees of freedom. 
This idea  is termed ``ER=EPR": entangled states (of the kind encountered in the EPR argument) can be represented by a spacetime that has a tiny `wormhole' (an Einstein-Rosen bridge ER) connecting the locations of the entangled states.

 The quanta $b_i, c_i$ in a Hawking emission process are in an entangled state which we may schematically represent as
 \be
|\psi\rangle={1\over \sqrt{2}}\left (|0\rangle_{b_i}|0\rangle_{c_i}+|1\rangle_{b_i}|1\rangle_{c_i}\right )
\label{oneq}
\ee

 One may therefore expect that one can replace this entangled pair by a thin wormhole connecting the locations of $b_i, c_i$. The set of  quanta $b_i$ is spread over a large region near infinity, but the ends of the wormholes at the locations of the $c_i$ can merge together to create what we normally think of as the  interior of the black hole. (Such a structure has been termed  a `squid'.) With this new geometry of spacetime, it is argued that one might be able to preserve the vacuum structure at the horizon while still getting purity of the emitted radiation \cite{maldasuss2}. 

\subsection{\label{secinter}Review of some earlier work}

Let us consider the duality pictured in fig.\ref{e2}(c),(d). The left and right CFTs are noninteracting. If each CFT state is dual to a bulk state, then we would get two noninteracting bulk configurations, as pictured in fig.\ref{e2}(d). The difference between the situations of fig.\ref{e2}(d) and fig.\ref{e2}(b) was investigated in \cite{holo}. The situation of fig.\ref{e2}(b) corresponds to a Hamiltonian with interaction between the left (L) and right (R) parts
\be
H_{connect}=H_L+H_R+H_{int}
\label{connect}
\ee
while that of fig.\ref{e2}(d) corresponds to a Hamiltonian with {\it no} interaction between these two parts
\be
H_{disconnect}=H_L+H_R
\label{disconnect}
\ee
It was then argued in \cite{holo} that while (\ref{disconnect}) is the correct dual of the two disconnected (but entangled) CFTs in fig.\ref{e2}(c), the situation can be {\it approximated} for some purposes by (\ref{connect}). Such approximate descriptions are the underlying idea of {\it fuzzball complementarity}. We will comment on the notion of fuzzball complementarity at the end, but for our discussion below we are interested in {\it exact} dualities only.  Note that standard AdS/CFT duality is exact, with all states and correlation functions in the CFT mapping exactly to states and correlations in the gravity dual. 

If the CFTs are disconnected, and each CFT state is dual to a gravity state, can we immediately conclude that a duality like fig.\ref{e2}(c,d) is correct and that of fig.\ref{e2}(a,b) is not? The answer is no, since the goal of the dual gravity description is to reproduce all the correlators of the CFT description. It could be the case that (\ref{disconnect}) is the correct gravity Hamiltonian, but adding the interaction as in (\ref{connect}) gives an {\it alternative} Hamiltonian, describing a connected spacetime. If this alternative description reproduces all the correlators of the CFT in an easy way, then the connected spacetime of fig.\ref{e2}(b) would be a useful dual description of the disconnected CFTs. In fact the computations of \cite{shenker} use the spacetime fig.\ref{e2}(b) to compute correlators, and find agreement with correlators in the CFT state of fig.\ref{e2}(a). Correlators in the right CFT  are reproduced by the gravity region R, correlators in the left CFT are reproduced by the gravity region L, while correlators with one leg on the left CFT and one on the right are reproduced with the help of regions F, P which  carry the information about the `entanglement' between the two CFTs. 

Marolf and Wall \cite{mw} have argued that the dual to entangled but disconnected CFTs should consist of differential superselection sectors; those where the gravity dual is connected like fig.\ref{e2}(b) and those where it is disconnected like fig.\ref{e2}(d). Recently, Avery and Chowdhury \cite{ac} have argued that the duality of fig.\ref{e2}(c,d) is correct while that of fig.\ref{e2}(a,b) is not. They argue that disconnected CFTs can have a dual gravity Hamiltonian of type (\ref{disconnect}), and not of type (\ref{connect}). They next argue that the gravity dual of fig.\ref{e2}(b) has correlators that have no dual interpretation in the CFTs, so the duality of fig.\ref{e2}(a,b) cannot be correct.\footnote{See \cite{mp1,ac2,amps2} for other discussions regarding the validity of  AdS/CFT duality in the black hole context.}

In the present paper we will use a somewhat different approach, by looking at the phenomenon of Hawking pair creation. This phenomenon will have different manifestations in the spacetimes fig.\ref{e2}(b) and fig.\ref{e2}(d). Comparing to what we expect in the CFT, we will get constraints on the picture fig.\ref{e2}(b).

 \section{\label{secthree}Challenges for the smooth horizon paradigm}
 
 We will now note four ways in which the Hawking pair creation process challenges the conjecture that the eternal hole is AdS is dual to an pair of noninteracting CFTs in the entangled thermofield double state (\ref{one}).

\subsection{\label{secthree:1}The `entanglement  current' at the horizon}

\begin{figure}[h]
\includegraphics[scale=.42]{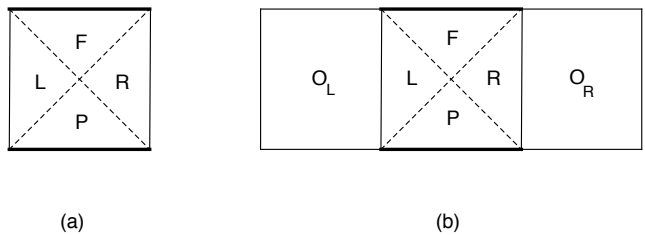}% Here is how to import EPS art
\caption{(a) The eternal hole in AdS, where Hawking radiation quanta reflect off the boundary and fall back into the hole. (b) We can extract the radiated quanta out to external regions $O_L, O_R$ by applying suitable operators at the boundary. \label{e1} }
\end{figure}

Consider a large black hole in AdS space (fig.\ref{e1}(a)). Quanta radiated from this hole reflect from the AdS boundary and fall back into the hole. Thus the entanglement of the hole with its radiation never becomes very large, and the  problem of growing entanglement does not become immediately manifest. 

To uncover the entanglement problem we couple the AdS boundary to an `outside' region, through operators that can be applied at this boundary. We can send particles into the AdS space from the AdS boundary, by applying a suitable operator at this boundary. Conversely, a particle that flies out from the hole to the  boundary of AdS can be `extracted' to the outside region, again by a suitable operator applied at the AdS boundary. We take the `outside' region to be asymptotically flat spacetime, where the extracted quanta can be analyzed at leisure far away from the hole. We can then ask for the entanglement of these extracted quanta with the AdS space containing the black hole. 

For the eternal hole in AdS we will have two `outside' regions, called $O_L, O_R$; we depict these in fig.\ref{e1}(b). We now argue as follows:

\b

\begin{figure}[h]
\includegraphics[scale=.62]{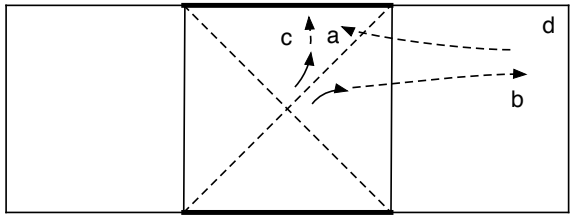}% Here is how to import EPS art
\caption{Hawking quanta b, c are produced at the horizon. The quanta b are extracted to the external region, while the c fall into the singularity. To restore the mass of the hole, we send in quanta a, which we can entangle beforehand with quanta d that stay in the exterior region.  \label{e3} }
\end{figure}

(i) The eternal hole appears to be a geometry where nothing changes with time. But in fact the state of the hole cannot be really time independent. If we wish to have a smooth horizon, then we have to choose the Hartle-Hawking vacuum at this horizon (as opposed to the Boulware vacuum). In the Hartle-Hawking vacuum we have the normal vacuum state in the local Kruskal coordinates straddling the horizon, and the usual Hawking computation gives the creations of entangled particle pairs. This pair creation happens at both the left and right horizons, as depicted in fig.\ref{e3}. 

(ii) We can extract the created particles to the outside regions by the application of suitable operators. Let us focus on the right (R) side. The extracted quantum $b$ will be entangled with a quanta $c$ falling into the hole, forming the state (\ref{one}).

(iii) The mass of the hole goes down when we extract the quantum $b$. To restore the mass, we throw in a quantum of energy $E\sim kT$. Let us call this quantum $a$. We could have entangled this quantum with another quantum $d$ before throwing it in;  we let $d$ be kept far away from the hole in the region $O_R$. We let the  entangled state of $a, d$  be
\be
|\chi\rangle={1\over \sqrt{2}}\left (|0\rangle_{a}|0\rangle_{d}+|1\rangle_{a}|1\rangle_{d}\right )
\label{onep}
\ee

(iv) At this stage we have restored the mass of the hole $M\rightarrow  M$, but the entanglement of the hole with the outside region $O=O_R+O_L$ has gone up by
\be
\delta S_{ent}=\ln 2
\label{increase}
\ee
This is of course just a restatement of how the eternal hole behaves. The hole is held in equilibrium with a heat bath at temperature $T$. But while this bath feeds the hole to keep the `mass flux' at zero, the `entanglement entropy flux' across the horizon is {\it not} zero. 

(v) We can keep this process going as long as we like. We can therefore make the entanglement $S_{ent}$ of the outside region $O$ with the AdS region as large as we want. But the AdS region has only a mass $M$ on each of the $L, R$ sides, and thus cannot have more than 
\be
Exp[S_{bek}(M)]\times Exp[S_{bek}(M)]=Exp[2S_{bek}(M)]
\label{ten}
\ee
states in the theory represented by the AdS space. By waiting till
\be
S_{ent}>2S_{bek}(M)
\label{contra}
\ee
we get a contradiction: the AdS region has only $Exp[2S_{bek}(M)]$ accessible states, but we have entangled this AdS region with the outside region $O$ in a fashion where the entanglement entropy  exceeds $2S_{bek}(M)$.

\b

\begin{figure}[h]
\includegraphics[scale=.42]{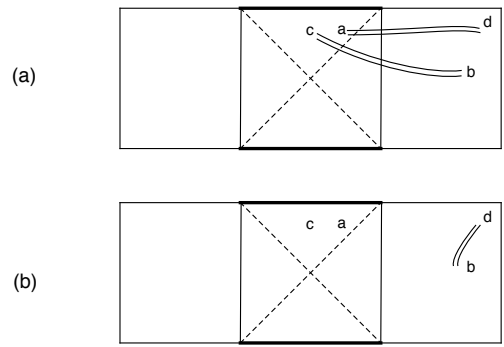}% Here is how to import EPS art
\caption{(a) In the picture of \cite{maldasuss2}, we can imagine thin wormholes connecting the entangled pairs b,c and a,d. (b) If the dynamics of wormholes is such that the entanglement moves to b,d, then we may be able to avoid the contradiction (\ref{contra}). \label{e4} }
\end{figure}

{\it Possible resolution:} \quad We should ask if it is possible to avoid this problem within the hypotheses that emerge in the approach (P1'). When the pair $b,c$ is created, we imagine a thin wormhole joins $b,c$, even when $b$ has been extracted to the region $O_R$. Similarly, there should be a thin wormhole connecting $a,d$ (fig.\ref{e4}(a)). 

Now suppose the ends of the wormholes at $c, a$ join up, leaving only one net wormhole joining $b,d$. Then the entanglement of the outside $O$ with the AdS region goes away, and we avoid a contradiction (fig.\ref{e4}(b)). 

But it is not clear what dynamics will lead to such a joining up of wormholes; showing such a dynamics (or some other resolution of the puzzle) is a challenge for the picture (P1').

\subsection{\label{secthree:2}Large entropy on `good slices'}

\begin{figure}[h]
\includegraphics[scale=.42]{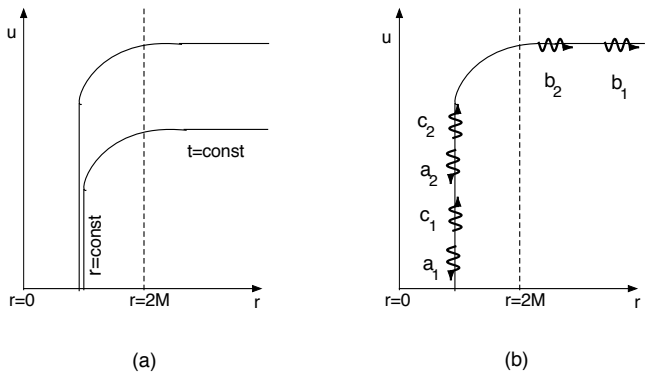}% Here is how to import EPS art
\caption{(a) A good slicing of the single sided Schwarzschild hole; the slices are depicted in Eddington-Finkelstein coordinates. (b) The entangled quanta $b_i, c_i$ on the good slice; the quanta $a_i$ on this slice are infalling quanta that maintain the mass of the hole. \label{e5} }
\end{figure}

Consider first the single sided black hole in asymptotically flat space, made by collapse of an initial shell of mass $M$. The Schwarzschild metric 
\be
ds^2=-(1-{2M\over r}) dt^2 + (1-{2M\over r})^{-1} dr^2+ r^2 d\Omega_2^2
\ee
 appears to be time independent, but such a time-independent description covers only the part $r>2M$ of the black hole spacetime. Spacelike slices are  $t=constant$ outside the hole, but $r=constant$ inside. Fig.\ref{e5}(a) depicts a `good slicing' in Eddington-Finkelstein coordinates. The part of the slice inside the hole asymptotes at early times to $r=M$. As we evolve along the slicing, this part becomes `longer' \cite{cern}. 
 
The negative energy quanta $c_i$ created in the Hawking process have wavelength $\sim M$ along this slice. We can maintain the mass of the hole by feeding in radiation at the temperature of the black hole; this gives positive energy quanta $a_i$ along this slice, again with wavelength $\sim M$. As we evolve to late times, the part of the slice at $r=M$ becomes very long, and collects a very large number of $c_i, a_i$ quanta (fig.\ref{e5}(b)). But these quanta contribute no net mass as seen from outside the sphere $r=2M$. This is because mass is measured by taking the inner product of the 4-momentum of the quantum with the Killing vector $\p_t$, and inside the horizon $\p_t$ points along the spatial direction along the slice. Thus the quanta $c_i$ with momentum in one direction along the slice contribute negative energy $(p_c, \p_t)<0$, while the quanta $a_i$ have momentum in the opposite direction and contribute positive energy $(p_a, \p_t)>0$.

By choosing different distributions for the  locations and spins of the $c_i, a_i$ we can  get a large number of states which all have approximately the  same energy $M$ as seen from outside the hole. We can make the number $N$ of these states as large as we wish by taking the part of the slice at $r=M$ to be sufficiently large. In particular we can make 
\be
\ln N>S_{bek}(M)
\label{large}
\ee
so that we have more states inside the hole than the Bekenstein entropy would allow. This is of course just another form of the information paradox. Similar constructions of states also arise in the puzzles known as the `bags of gold' problem \cite{gold}, and the `monsters' construction \cite{monsters}.  

 \begin{figure}[h]
\includegraphics[scale=.52]{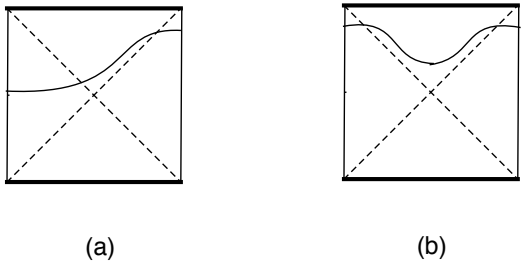}% Here is how to import EPS art
\caption{(a) A good slice analogous to the slice of fig.\ref{e5}(a); such a  slice can hold more entropy than the Bekenstein entropy of the hole. (b) We can extend such slices to both sides of the eternal hole, compounding  the  problem. \label{e6} }
\end{figure}

But we can now consider the same problem for the eternal black hole in AdS. The analogue of the slice in fig.\ref{e5} is shown in fig.\ref{e6}(a). One can extend the slice to both sides of the eternal hole; this gives the slice in fig.\ref{e6}(b). 

In the dual CFT we have only $Exp[2S_{bek}(M)]$ states for the given energy in the CFT (eq.(\ref{ten})). But the slices in fig.\ref{e6}(a) or fig.\ref{e6}(b)  can carry  
a number of states 
\be
N>Exp[2S_{bek}(M)]
\ee
This is a contradiction: why are there more states in the bulk than in the CFT?

\b

{\it Possible resolutions:} \quad One might try to argue that the states on this good slice in the bulk are somehow pure gauge states of the bulk theory, which need not find a dual description in the CFT. But there is no clear reason for this: on a good slice the quanta $c_i, a_i$ are normal local particle modes, and the entropy of such states can be computed just the same way one would compute the entropy of a gas of low energy quanta in a gas.\footnote{Hartman and Maldacena \cite{hartmal} have developed a `tensor network' description of processes which could model unitary evolution of a black hole. However, it is not known how to map such a model to the gravity description being considered here.}

We should also point out what would {\it not} be a resolution to our puzzle. We cannot simply claim that the good slices of fig.\ref{e6} should not exist, or that states on such slices should not be `normal' low energy states. . Such an argument would be `begging the question'. It is already  clear that if we disallow such slices (or states on them) then we can avoid the entropy problem mentioned above \cite{stretch1, lm4, stretch2}\footnote{See also \cite{giddingsslice}.}. What one needs is a {\it mechanism} to destroy such states on the good slice when the slice becomes too long. Finding such a mechanism is the crucial issue; the region around the slice is a region of gentle curvature, and semiclassical physics should be valid in such a situation unless we pinpoint an effect that would violate it.  In the fuzzball paradigm, a specific mechanism has indeed been proposed: tunneling into the large class of fuzzball states \cite{tunnel} (see below). This mechanism is nontrivial, since it does not operate in a theory without fuzzballs.  For example in canonically quantized 3+1 gravity where we do not have fuzzballs,   we would indeed get the above described states with large degeneracy (\ref{large}) on the good slice.

\subsection{\label{secthree:3}The `left-right problem'}

Suppose we throw in a particle $p$ from the external region $O_R$ onto the eternal hole. In the CFT description, $p$ excites the right CFT. Since the CFTs are disconnected, we expect that the information of $p$ will ultimately be radiated back into the region $O_R$, and not into the region $O_L$. But from the bulk perspective, the left and right sides of the eternal hole are connected, and we can arrange things so that it appears plausible that the information of $p$ emerges into the {\it left} exterior region $O_L$. This appears to be a conflict caused by the connectedness of the eternal hole spacetime. We now discuss this problem in more detail.

 \begin{figure}[h]
\includegraphics[scale=.54]{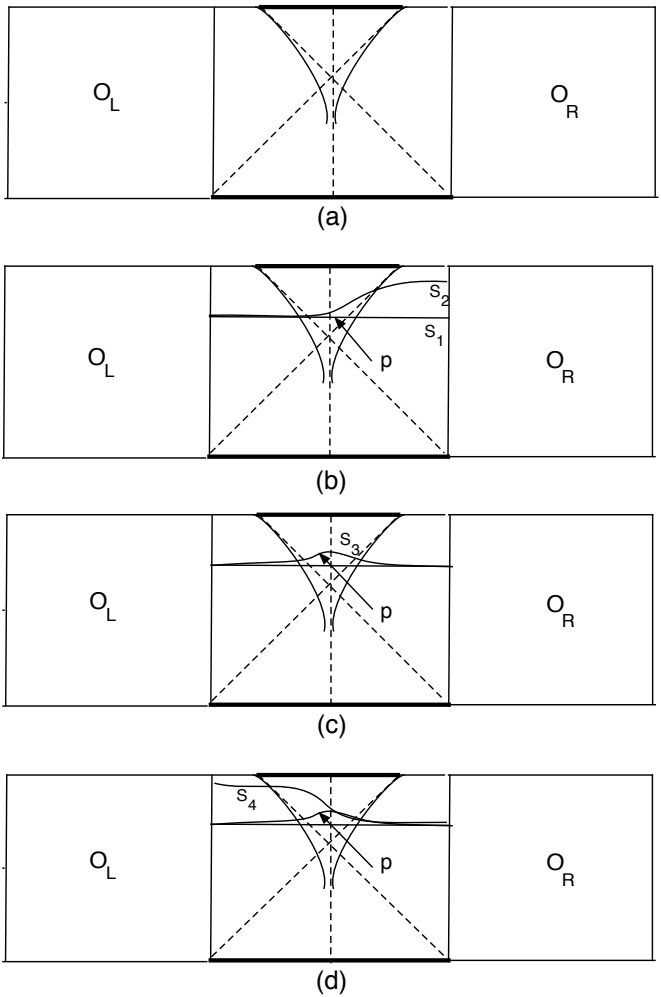}% Here is how to import EPS art
\caption{(a) The 2-sided hole in AdS, which is allowed to evaporate on both left and right sides by extracting quanta to the exterior regions $O_L, O_R$. (b) The infalling quantum $p$ is caught on slice $S_1$; this slice is evolved up in time to the right (slice $S_2$), so that we may expect the information in $p$ to emerge in $O_R$. (c)  The slice $S_1$ is evolved to a slice $S_3$, so that $p$ reaches to the left of the central line (this evolution happens in crossing time, so no significant information is expected to be emitted) (d) The slice $S_3$ is evolved up in time to the left (giving slice $S_4$); in this situation we may expect the information of $p$ to emerge on the left. \label{e7} }
\end{figure}

Consider again the set-up of fig.\ref{e1}(b), where we have coupled the eternal hole to external regions $O_L, O_R$  on the left and right sides. We throw in a particle $p$ from $O_R$ into the AdS space. This particle will fall through the horizon on the right side. Now we let the black hole evaporate, by extracting to the region $O_L$ the Hawking particles that come to the AdS boundary on the L side and extracting to the region $O_R$ the Hawking particles that come to the AdS boundary on the R side.

We expect the evaporation to be unitary, so the information of the particle $p$ should be encoded in the Hawking radiation. But will this information be encoded in the radiation on the L side, the R side, or shared between both?

Consider the CFT description. The particle $p$ has fallen into the R side of the hole, and is therefore an excitation of the R CFT. Since the CFTs are non-interacting, we would expect that the information should be encoded in the radiation emerging on the R side only. In the bulk, the particle $p$ is on the slice $S_1$ (fig.\ref{e7}(a)). On this slice, $p$ is on the right side of the vertical dividing line drawn through middle of the eternal black hole  diagram. Next, we perform the evolution using the slices shown in fig.\ref{e7}(b), which correspond to  advancing  the Schwarzschild time on the R side. Radiation emerges to the right, and we expect that this radiation encodes the information in the particle $p$. 

But in a theory of gravity we can advance our spacelike slices in any manner that we wish (at least in a region where the semiclassical approximation is valid). Thus consider the alternative evolution shown in fig.\ref{e7}(c). This evolution from slice $S_1$ to slice $S_3$ sends $p$ {\it across} the vertical dividing line drawn through middle of the eternal black hole  diagram. This evolution happens over a time of order the crossing time, so no significant information about $p$ is emitted during the evolution from $S_1$ to $S_3$. Next, we  perform the evolution using the slices shown in fig.\ref{e7}(d), which correspond to  advancing  the Schwarzschild time on the L side. In this evolution particle pairs are created at the left horizon, and  we may expect that the information will come out on the {\it left}  side. 

So this is the puzzle: does the information of the particle $p$ come out on the R side or on the L side (or a little on each side)?

\b

 \begin{figure}[h]
\includegraphics[scale=.54]{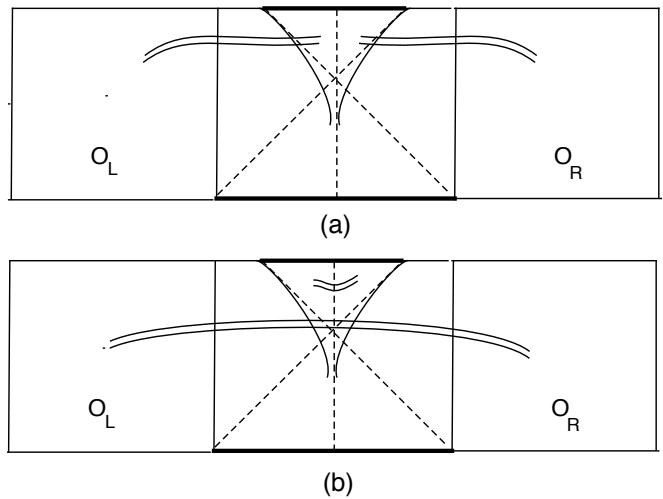}% Here is how to import EPS art
\caption{(a) Pair creation at each horizon gives wormholes connecting the interior regions to their corresponding exterior regions. (b) If the evolution of wormholes leaves $O_L$ connected to $O_R$, then it may not be possible to sharply localize information on the left and right sides separately. \label{e8} }
\end{figure}

{\it Possible resolutions:}  (i)  From the CFT description it appears hard to avoid the conclusion that the information of $p$ should be encoded in the right CFT alone. But once we extract quanta to the outside region $O_R$, then the situation is less clear. We have seen that the radiation of Hawking pairs $b,c$ can cause wormholes to connect the right half of the eternal hole to $O_R$, and the left half of the eternal hole to $O_L$. But the further dynamics of these wormholes can lead to the situation depicted in fig.\ref{e8}(b), where wormholes connect the exterior region $O_L$ directly to the exterior region $O_R$. It would then be hard to localize information in $O_R$ or $O_L$ alone, since these regions are not really disjoint; all one can say is that the information is distributed in the entire exterior region $O_L+O_R$. This situation would bypass the  left-right puzzle by blurring the distinction between the left and right exterior regions. But it is then a challenge to show that the situation of fig.\ref{e8}(b) indeed arises from the detailed dynamics of wormholes.

\b

(ii) Alternatively, we can give up the idea that the exterior regions $O_L$ and $O_R$ are connected as in fig.\ref{e8}(b), but ask that the dynamics of information retrieval be such that we resolve the left-right problem. The particle $p$ has fallen in from the right, so it has a momentum pointing to the left. We may conjecture that if $p$ has a momentum to the left, then its information will come out on the right, regardless of whether the {\it position} of $p$ is to the left or right of the central vertical line in fig.\ref{e7}.
If the physics of information recovery satisfied this property, then it might seem that we avoid the left-right problem.

But there are difficulties in implementing such a solution. Consider a single sided black hole, made from collapse of a shell. Inside the horizon we have particles moving to the left (infalling particles) and particle moving to the right (for example negative energy partners $c$ of the Hawking quanta $b$). Since there is only one exterior region, we expect that the information of {\it both} left and right moving particles, emerges on the {\it same} side. How would we reconcile this fact with the  requirement that in the eternal hole particles with  different momenta  radiate information to different sides?

\b

We have thus seen three potential problems (listed in sections \ref{secthree:1}, \ref{secthree:2}, \ref{secthree:3}) for the conjecture that entangled but disconnected CFTs should have a gravity dual that is the connected spacetime of the eternal hole. We now mention how these problems might be addressed in the fuzzball paradigm.

 \section{\label{secfour}Contrasting the approach (P1') with the fuzzball approach (P2')}

In the fuzzball approach (P2'), the CFT is dual to a set of microstates that end before a horizon is reached. Thus if we take two CFTs in an entangled state, we expect to get just two spacetime solutions that are entangled with each other (fig.\ref{e2}(c,d)). Let us now  examine how the above issues (A),(B),(C) would be treated in the fuzzball scenario.

\subsection{\label{secfour:1}Entanglement current}

In the fuzzball picture we do not create entangled pairs from the vacuum - the radiation arises from the fuzzball surface just like it would from a piece of coal \cite{radiate}. Thus this issue does not arise at all. 

\subsection{\label{secfour:2}Large entropy on good slices}

In the fuzzball picture it is conjectured that the good slices cannot be stretched in the manner where they can carry a large entropy. Before the slice evolves to this stage, the geometry gets altered by tunneling into the large space of fuzzball states \cite{tunnel,rate},\cite{beyond}. Thus this issue does not arise either.

\subsection{\label{secfour:3}The Left-Right problem}

This problem suggests that the the geometry of the eternal hole is itself incorrect. That is, if we start with the state given at the $t=0$ slice of this geometry (the slice $S_1$ in fig.\ref{e7}(a)), then even over the crossing timescale $M$ we do not obtain the traditional geometry of the eternal hole. Instead, the spacetime gets corrupted almost immediately.  Let us see how this might happen.

In the fuzzball proposal it is argued that tunneling into fuzzball states alters the semiclassical spacetime geometry in a  way that would not happen in theories without fuzzballs \cite{tunnel}. The rapid rate of tunneling stems from the fact that while the amplitude for tunneling into a fuzzball state may be   small
\be
{\cal A} \sim e^{-\alpha GM^2}, ~~~\alpha\sim O(1)
\label{el}
\ee
the number of states to which one can tunnel is large
\be
{\cal N} \sim e^{S_{bek}}\sim e^{4\pi G M^2}
\label{tw}
\ee
Thus it is possible that the smallness of the tunneling probability $|{\cal A}|^2$ cancels against the largeness of ${\cal N}$, and the semiclassical geometry gets destroyed as a consequence. 

A priori, there are different possible values for the timescale $t_{tunnel}$ over which this tunneling might happen:

(i) $t_{tunnel}\ll t_{evap}$. In \cite{rate} an argument was given for this bound on the tunneling time; such an estimate resolves the information paradox.

(ii) $t_{tunnel}\lesssim t_{scrambling}$. This estimate was suggested by requirements of fuzzball complementarity \cite{mt2}.

(iii) $t_{tunnel}\sim t_{crossing}$. The crossing time scale is another relevant scale in the black hole problem.

(iv) $t_{tunnel}\ll t_{crossing}$. If tunneling happens this rapidly, then we would not get the traditional eternal hole geometry at all. This is the case because a slice like $S_3$ in fig.\ref{e7}(c) has a part inside the horizon, and such regions would be immediately destabilized by tunneling into fuzzballs.  If the eternal hole spacetime is not a good semiclassical geometry, then we avoid the left-right problem which results from different semiclassical foliations of this spaectime.

This rapid tunnelling conjectured in (iv) is possible because we have a competition between two large exponentials in  (\ref{el}), (\ref{tw}). If we compress matter to a radius that is smaller than the horizon, then it appears plausible that we would get to a situation where the degeneracy of states overwhelms the smallness of the tunneling probability
\be
{\cal N} \gg |{\cal A}|^{-2}
\label{condition}
\ee
so the configuration tunnels extremely rapidly into fuzzballs, destroying the semiclassical geometry. Let us discuss this conjecture in more detail.  

\section{\label{conjecture}The conjecture of `quick tunneling'}

\begin{figure}[h]
\includegraphics[scale=.54]{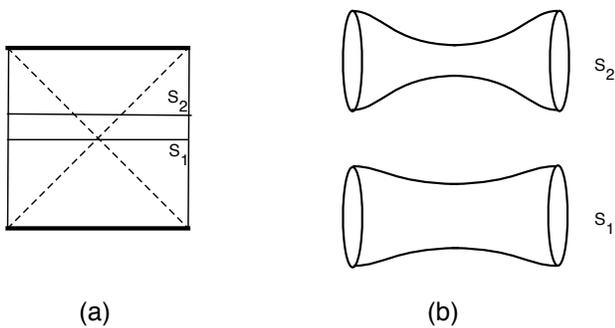}% Here is how to import EPS art
\caption{(a) The eternal hole, where we draw two spacelike slices $S_1, S_2$. (b) The slices $S_1, S_2$ have the geometry of Einstein-Rosen bridges; we have depicted the angular direction along with the radial direction seen in (a). The slice $S_1$ has a minimum radius $r_{min}=2M$, while $S_2$ has a minimum radius $r_{min}<2M$.  \label{e12} }
\end{figure}

 Fig.\ref{e12}(a) depicts two slices $S_1$ and $S_2$ which give Einstein-Rosen bridges with different values of the `minimum radius'. Fig.\ref{e12}(b) depicts these two bridges. Classically, one would think that the bridge for the slice $S_1$ (with momenta $\pi_{ij}=0$)  provides a good initial condition for evolving the spacetime, yielding the eternal hole spacetime of fig.\ref{e12}(a). But the later slices in the geometry, like slice $S_2$, are Einstein-Rosen bridges with a minimum radius  $r_{min}<2M$ (fig.\ref{e12}(b)).\footnote{For ease of discussion, we use $r=2M$ as the horizon radius of the hole; as a matter of fact the horizon radius is a function of the mass and the AdS radius since this eternal hole is the large black hole in AdS space.} We conjecture that such spacelike slices do not exist in a good geometry: because of (\ref{condition}), they are destroyed by tunneling in a time that is short compared to the curvature length scale.

\begin{figure}[h]
\includegraphics[scale=.42]{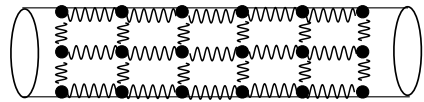}% Here is how to import EPS art
\caption{\label{e10} The slice $S_1$ of fig.\ref{e12}(a); the horizontal direction is the radial coordinate, while the circular direction is a compact direction of spacetime (the angular direction depicted in fig.\ref{e12}(b) has been suppressed). The spatial slice has been approximated by a lattice of points on which the variables $g_i$ live; the gradient term $\nabla g$ is represented by springs joining the lattice points.}
\end{figure}

To understand this conjecture in more detail, we recall the picture of spacelike slices discussed in \cite{thermo}. Let us assume that our spacetime has a compact circle. In the fuzzball construction,  the  shrinking  of this circle to zero size causes the spacetime to end outside the horizon. Consider the spacelike slice $S_1$ of fig.\ref{e12}(a). In fig.\ref{e10} we depict this spacelike slice as stretching along the horizontal direction, and the compact circle over each point of this slice gives the depicted cylinder. We have regulated the gravity theory by taking a lattice of points on this slice, represented by the black dots; on each lattice site we have field variables $\{ g_i\}$ representing the gravitational and other fields of the theory. These variables $g_i$  have kinetic terms $\nabla g$ which we have indicated by springs joining the sites.\footnote{We can get an approximate model of a  free quantum field theory by taking a set of masses joined by springs; we have used a crude model of this type to give a qualitative depiction of the point we wish to make.} The wavefunction on this slice is $\Psi[\{g_i\}]$, with $i$ running over  the lattice sites on the slice. 
The left half of this cylinder lies in the left half  of the eternal hole shown in fig.\ref{e12}(a), while the right half lies in the right half of the eternal hole. The Hamiltonian given by the links in fig.\ref{e10} is of the form (\ref{connect}): $H_{connect}=H_L+H_R+H_{int}$, since we have bonds between the left and right sides of the cylinder. 

\begin{figure}[h]
\includegraphics[scale=.42]{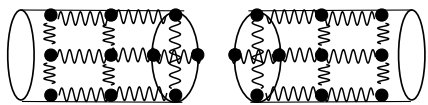}% Here is how to import EPS art
\caption{\label{e11} The same lattice of points as in fig.\ref{e10}, but with different springs connecting them. We can expand the same  state $\Psi[g_i]$ either as eigenfunctions of the Hamiltonian given by the links in fig.\ref{e10}, or by the links depicted above. In the case above, the inner ends of the cylinders have points identified with their diametrically opposite points, causing each of the cylinders to end in a cross-cap. }
\end{figure}

In fig.\ref{e11}, we take the same cylinder as in fig.\ref{e10}, with the same variables $g_i$ on the same lattice sites. But we imagine these sites to be joined by different links.  The points on the cylinder have been divided into those on the left side and those on the right side.  At the end of each cylinder, a lattice site is linked to its diametrically opposite site. Thus the ends of the cylinders are `sewn-up', to produce a `capped' geometry.\footnote{In this simple example we get a `cross-cap', but in the full fuzzball geometry we expect the cap to be made of  KK-monopoles.} This set of `springs' on the links gives a {\it different} Hamiltonian  that acts on the {\it same} state $\Psi[\{g_i\}]$. Since the left and right sides of the cylinder are now disconnected, we have a  Hamiltonian of the form (\ref{disconnect}): $H_{disconnect}=H_L+H_R$.

We can now state the `quick tunneling' conjecture. Suppose we set up our initial data to have the  above discussed wavefunctional $\Psi[\{g_i\}]$ on the slice $S_1$ of fig.\ref{e12}(a).  If there were no fuzzballs, we might expect that we can expand $\Psi[\{g_i\}]$ in eigenstates of  the Hamiltonian $H_{connect}$, and  evolve to get the eternal black hole spacetime of fig.\ref{e12}(a). But this evolution would lead us to slices like $S_2$, where the minimum radius is $r_{min}<2M$. In a theory with fuzzballs, the conjecture of `quick tunneling' would say that there are no good semiclassical geometries where the minimum radius can reach a value $r_{min}<2M$; such geometries are destroyed  by tunneling into the vast space of fuzzball states.\footnote{We can restate this in the following words. Consider the wavefunctional $\Psi[g]$ on the state of all metrics $g$. The existence of fuzzball states means that there are a large number of directions in the space of metrics $g$.  Suppose we wish to make a wavepacket that is  peaked not around fuzzballs, but around a smooth slice. If this slice has   $r<2M$ at some point, then the wavefuntion can spread to the fuzzball states. To make the wavefunction peaked on the semiclassical slice, we need to make the wavefunction fall off rapidly in all directions leading towards the fuzzball states. This fall-off provides a contribution to the energy carried by the wavefunctional $\Psi$, and prevents  $\Psi$ from being peaked in energy around $\sim M$. So we do not get a `good' decomposition of $\Psi$ into eigenstates of $H_{connect}$.} What we should do instead is expand $\Psi[\{g_i\}]$ in eigenstates of the Hamiltonian $H_{disconnect}$, and evolve; this will give rise to the disconnected but entangled spacetimes  of fig.\ref{e1}(d). In this latter situation, we have a good semiclassical geometry up till the vicinity of the horizon, and then a quantum mess of `caps' as we approach the location where the horizon would have been. 

To summarize, the `quick tunneling' conjecture  says that we cannot confine matter with a mass $M$ within a radius $r<2M$. If we do try to construct an initial state where the mass $M$ is so confined on a smooth slice, then the subsequent evolution quickly destroys the semiclassical nature of the slice by a tunneling into fuzzballs (which have structure at $r\gtrsim 2M$ and no `interior'). Thus initial conditions with matter confined to $r<2M$ do not lead to good initial date for a semiclassical evolution. This `quick tunneling' conjecture would resolve our left-right problem, since we cannot perform the evolution in fig.\ref{e7}(c) which takes a particle from the left side of the geometry to the right.

\section{Discussion}

In this paper we have shown certain challenges faced by the idea \cite{eternal} that the dual of two entangled but disconnected CFTs is the connected spacetime of the eternal hole (fig.\ref{e1}(a,b)). We have also noted some possibilities on how these challenges might be addressed in the wormhole picture proposed in \cite{maldasuss2}. But the challenges impose several constraints on the behavior of the wormholes, and at present it is not clear how these required properties of wormholes will emerge. 

\subsection{The constraints from Hawking radiation}

The duality of disconnected but entangled CFTs to the eternal hole has also been questioned in \cite{mw, ac}. Our arguments have been a little different, as they  are all based on the nature of Hawking radiation. We have noted that if the eternal hole spacetime is assumed to have smooth future horizons, then we cannot avoid Hawking radiation from these horizons. Though we can keep  the mass of the hole constant by allowing quanta to fall into the hole, the entanglement of the hole with the outside nevertheless increases (eq.(\ref{increase})). We get all the usual information issues created by this rising entanglement, and these issues have led to our challenges A-C described in section \ref{secthree}.

We have also noted how the challenges we raised can be addressed in the fuzzball picture, where the entangled but disconnected CFTs are dual to a pair of entangled but disconnected spacetimes (fig.\ref{e1}(c,d)). Two of the challenges were automatically met by the fuzzball picture while the third (the left-right problem) suggested the `quick tunneling' conjecture where the  tunneling into fuzzballs \cite{tunnel} is rapid enough to disallow semiclassical behavior for any slice where a mass $M$ has been confined to a radius $r<2M$. 

\subsection{Fuzzball complementarity}

Even though we have argued that the eternal hole spacetime fig.\ref{e1}(b) does not arise as the dual of two disconnected CFTs, the idea of fuzzball complementarity nevertheless indicates the role of this spacetime as a tool to obtain an {\it approximate}  description of processes in the disconnected spacetimes fig.\ref{e1}(d). The approximation is valid for impacts on the fuzzball surface that arise from quanta with $E\gg T$; i.e., the energy of the infalling quantum is much larger than the temperature of the hole. In this situation the impact creates a large number of {\it new} degrees of freedom. In fact the number of states $N_f$ after the impact is related to the number $N_i$ before the impact through \cite{mt2}
\be
{N_f\over N_i}={Exp[S_{bek}[M+E]]\over Exp[S_{bek}[M]]}\approx  e^{E\over T}
\label{nfniq}
\ee
Thus ${N_f\over N_i}\gg 1$ for $E\gg T$.  The dynamics of these newly accessed degrees of freedom is a `collective dynamics' of the fuzzball, and it is this dynamics which is conjectured to be reproduced by the eternal hole spacetime. Note that the newly accessed degrees of freedom are not entangled with anything else; in particular they are not entangled with radiation that may have escaped earlier from the hole. We thus note that while   the AMPS argument \cite{amps} rules out traditional complementarity, it does not rule out fuzzball complementarity \cite{mt2}.  

If the conjecture of fuzzball complementarity is true, then we get the eternal hole spacetime as an approximate tool for high impact processes. But note that this is a very different from having the eternal hole as the true dual of two entangled CFTs. In the latter case, the effect of the entanglement in (\ref{one}) is seen only at energies $E\lesssim T$, since for $E\gg T$ there is no appreciable entanglement between the two CFTs. With fuzzball complementarity, on the other hand, the effective spacetime of the eternal hole emerges in the opposite limit $E\gg T$. 

\subsection{Summary}

We face an apparent paradox in fig.\ref{e2}. Suppose we start from the eternal black hole spacetime in fig.\ref{e2}(b). Then AdS/CFT suggests that the dual is two entangled but  disconnected CFTs, since there are two disconnected AdS boundaries, (fig.\ref{e2}(a)).
But if we start with two disconnected CFTs (fig.\ref{e2}(c)), then each CFT is dual its own spacetime, so we expect two entangled but disconnected spacetimes (fig.\ref{e2}(d)).

There are three possible resolutions of this conundrum: (i) The duality of fig.\ref{e2}(a,b) is correct (ii) The duality of fig.\ref{e2}(c,d) is correct (iii) The descriptions of fig.\ref{e2}(b) and fig.\ref{e2}(d) are equivalent, so both dualities are correct. 

We have argued for possibility (ii). We have noted  that the process of Hawking radiation places severe constraints on  the duality fig.\ref{e2}(a,b), though we have tried to list possible escape routes using the idea of wormholes developed in \cite{maldasuss2}. We have noted how these constraints are bypassed in  the duality fig.\ref{e2}(c,d), which is based on the idea of fuzzball states. In the process we have conjectured that the eternal black hole spacetime of fig.\ref{e2}(b) is itself incorrect, since it has regions where the geometry appears to have a mass $M$ localized within a radius $r<2M$; such situations are disallowed due to the process of tunneling into fuzzballs. 

\begin{acknowledgments}
I thank Ramy Brustein, Borun Chowdhury, Juan Maldacena, Ashoke Sen, S. Shenker,  David Turton  and H. Verlinde for discussions.
This work was supported in part by DOE grant DE-FG02-91ER-40690.
\end{acknowledgments}

\nocite{*}

%\bibliography{counting}% Produces the bibliography via BibTeX.

\end{document}